\documentclass{iau_FM}
\usepackage{graphicx}

\title[Rotation curves of disk galaxies and GR] 
{Rotation curves of disk galaxies and\\ General Relativity}

\author[Luca Ciotti]   
{Luca Ciotti}

\affiliation{Dept. of Physics and Astronomy, University of Bologna, \\ 
via Piero Gobetti 93/2, I-40129, Bologna, Italy \\ email: {\tt luca.ciotti@unibo.it} \\[\affilskip]
}

\pubyear{2024}
\jname{Astronomy in Focus, Focus Meeting 9} 
\editors{Diana M.~Worrall, ed.}
\begin{document}

\maketitle

\begin{abstract}

It has been proposed that the flat rotation curves observed at large radii in disk galaxies can be interpreted as an effect of  General Relativity (GR) instead of the presence of dark matter (DM) halos in Newtonian gravity. In Ciotti (2022) the problem is rigorously explored in the special setting of the weak-field, low-velocity gravitomagnetic limit of GR. The rotation curves are obtained for purely baryonic disk models with realistic density profiles, and compared with the predictions of Newtonian gravity for the same disks, in absence of DM. The rotation curves are indistinguishable, with percentual GR corrections at all radii of the order  of $\approx 10^{-6}$ or less, so that DM halos are required in gravitomagnetism as in Newtonian gravity. From a more general point of view, a list of the most urgent problems that must be addressed by any proposed GR-based alternative to the existence of DM, is given.

\end{abstract}

\firstsection 
\section{Introduction}

Recently it has been suggested that flat rotation curves in disk galaxies could be a peculiar effect of General Relativity (hereafter GR) in rotating systems, without the need to invoke the presence of DM halos (as a very incomplete list, see e.g. Cooperstock \& Tieu 2007; Balasin \& Grumiller 2008; Crosta et al. 2020; Ludwig 2021,2022; Astesiano \& Ruggiero 2022; Beordo et al. 2024; Galoppo \& Wiltshire 2024; Galoppo et al. 2024, and references therein). If confirmed, the suggestion would be most surprising because disk galaxies are stellar systems in the low velocity, weak field regime of GR, with $v/c\approx 10^{-3}$, so that the common expectation is of GR corrections to Newtonian predictions of the order of $(v/c)^2\approx 10^{-6}$, enormously smaller than the {\it empirically required} effects of the order of (say) $\approx 500\%$, commonly attributed to the presence of massive DM halos.

In Ciotti (2022) the expected rotation curves of axisymmetric disk galaxy models in absence of DM, and with total baryonic mass, scale length, and density profile similar to those observed in real disk galaxies, are rigorously obtained in Newtonian gravity and in the well known low velocity, weak field gravitomagnetic limit of GR (Landau \& Lifshitz 1971, see also Mashhoon et al. 1999; Ruggiero \& Tartaglia 2002; Poisson \& Will 2014). In particular, the general integral expressions of the gravitomagnetic fields associated with the disk mass current, are given both in terms of elliptic integrals and Bessel functions, and discussed from the mathematical and physical point of view. For the case of the rotation curve of generic razor-thin disk of total mass $M_{\rm disk}$ and scale-lenght $R_{\rm disk}$, supported by circular orbits, and no DM, a regular series expansion of the {\it self-consistent} GR rotational velocity profile $v_{\rm GR}(R)$ in terms of the {\it naturally occurring} expansion parameter $\epsilon = G M_{disk}/(R_{\rm disk} c^2)\simeq 10^{-6}$ is formally derived
\begin{equation}
v_{\rm GR}(R)\sim v_{\rm N}(R)+\epsilon\,v_1(R),
\end{equation}
where $v_{\rm N}(R)$ is the circular velocity profile of the disk without DM in Newtonian gravity (e.g. Ciotti 2021), and $v_1(R)$ is the self-consistent gravitomagnetic correction. It is found that $v_1$ is already significantly smaller than $v_{\rm N}$ at all radii, so that after multiplication by $\epsilon\simeq 10^{-6}$ (that we stress is not introduced ``ad hoc'' but appears naturally in the mathematical treatment of the problem), we obtain {\it essentially identical} Newtonian gravity and gravitomagnetic rotation curves at all radii: in particular, in the outer regions of the disk, the GR rotation curves are Keplerian. We notice that, as explained by the gravitomagnetic Biot-Savart analogue, while in the inner regions of the disk $v_1 >0$, in the external regions $v_{1}<0$, i.e., if any, the correction effects actually tend to {\it decrease} the rotational velocity below the Keplerian profile! All these results are then confirmed repeating the analysis in the different family of disks, described by the Kuzmin (1956) profile. The assumption of razor-thin disks is then relaxed, and the gravitomagnetic Jeans equations for collisionless axisymmetric systems supported by two-integrals phase-space distribution functions are derived. The case of the Miyamoto-Nagai (1975) disk is studied in detail, again fully confirming the results for razor-thin disks: no detectable differences between Newtonian gravity without DM and GR rotation curves are predicted. In conclusion, Ciotti (2022) seems to excludes the possibility that gravitomagnetic GR effects can compensate by any amount the Keplerian fall of the rotational  velocity that would characterize disk galaxies at large galactocentric distances in absence of DM halos, and produce the observed flat profiles: DM is required by GR in disk galaxies exactly as in Newtonian gravity. 

\section{Problems to be addressed by GR-based alternatives to DM}

As anticipated above, gravitomagnetism is only one out of many scenarios based on GR (or on some of its variants) to avoid the introduction of DM in order to explain observations. These proposals are quite sophisticated, and usually involve a non trivial amount of advanced mathematical manipulations, that unfortunately tend to obscure the {\it physical} origin of the claimed very large contributions to the observed velocities (``corrections'' larger than the velocities predicted by Newtonian gravity in absence of DM!). For this reason we present here a (partial) list of key astrophysical problems that should be convincingly addressed by any proposed attempt to replace DM halos by GR effects in (weak-field) astronomical systems (see also Ciotti 2023): the first of the problems has already been mentioned above.

\begin{itemize}

\item
Stellar velocities in massive galaxies are $v\approx 300$ km/s, and proportionally less in lower mass stellar systems (some of them requiring in Newtonian gravity even larger values of DM-to-baryon ratios than disk galaxies!), with  expected  corrections of $(v/c)^2\approx 10^{-6}$. The needed GR corrections of the Newtonian predictions instead should be of the same order of the observed velocities: a physically clear explanation of this huge and unexpected effect is needed, hopefully presented with a simple but robust order-of-magnitude estimate, involving the observed galactic properties.
\item
Several GR-based conceptual frameworks have been proposed, from gravitomagnetism, to geometric dragging, to GR effects of boundary conditions/vacuum solutions, to retarded potential effects due to unsteady accretion of gas on galaxies, to gravitomagnetic dipole effects produced by pairs of rotating BHs, and so on. To the Astronomer, it looks there are {\it too many} GR proposed solutions: are they different manifestation of the same phenomenon? are they physically different? are they compatible?
\item
Exponential disks in Newtonian gravity produce reasonably flat rotation curves in the radial range from 1 to 3 scale-lengths, even in absence of a DM halo. DM is only required by HI rotation curves in the external regions, well beyond the edge of the bright optical part of the galaxy. A GR model predicting a flat rotation curve inside the stellar disk is just in accordance with Newtonian gravity.
\item
If DM is mimicked by GR effects due to rotation (such as in gravitomagnetism), why massive DM halos are required also in Clusters of Galaxies and Elliptical Galaxies, systems that often show a {\it very low} rotational support? It should be recalled that rotation curves of disk galaxies are not the only (and not the most important) phenomenological indication of the possible presence of DM halos.
\item
For galaxies/clusters with gravitational lensing (a GR weak field prediction), DM halos are inferred, with properties in remarkable agreement with the predictions based on stellar dynamics and/or hydrostatic equilibrium of hot, X-ray emitting gaseous coronae. Should we conclude that also gravitational lensing is affected by (unexpected) GR corrections?
\item
Dwarf spheroidal galaxies and ultra-faint galaxies are low-mass stellar systems, with very small values of $v/c$ but with inferred very high DM-to-baryon ratios, even larger than in disk galaxies. Why GR effects in these very weak-field systems are proportionally {\it more important} than in more massive galaxies? 
\item
In Globular Clusters, some of them in the same shape/velocity dispersion/mass range of galaxies in previous point, DM {\it is not} required: why GR should behave so differently in stellar systems of similar mass? 
\item
Cosmological simulations with DM and initial conditions derived from the observed CMB appear to be highly successful in reproducing the growth of the large scale structure of the Universe. What about the cosmological predictions of GR in absence of DM?
\item
Disk galaxies in absence of DM halos are unstable in Newtonian gravity (Ostriker \& Peebles 1973): can pure GR stabilize disk galaxies? 
\item
Finally, it would be of the utmost importance to have clear and well defined mathematical {\it procedures} to be applied to the {\it observed baryonic components} of galaxies to unambiguously predict the expected GR effects in the proposed scenario, to be compared with the observed kinematical fields.

\end{itemize}

\end{document}